\documentclass[aps,prd,twocolumn,10pt,superscriptaddress,nofootinbib,nobibnotes,longbibliography]{revtex4-1}
\usepackage{graphicx}
\usepackage{color}
\usepackage{xcolor}
\usepackage{dcolumn}
\usepackage{bm}
\usepackage{slashed}
\usepackage{amsmath}
\usepackage{latexsym}
\usepackage{amssymb}
\usepackage{mathrsfs}
\usepackage{amsfonts}
\usepackage{hyperref}
\usepackage{xspace}
\allowdisplaybreaks

\begin{document}
\title{A cosmological intercept tension in local distance ladders}

\author{Haotian Dong}
\affiliation{School of Physics and Astronomy, Beijing Normal University, Beijing 100875, China}
\affiliation{National Centre for Nuclear Research, Pasteura 7, PL-02-093 Warsaw, Poland}

\author{Marek Biesiada}
\affiliation{National Centre for Nuclear Research, Pasteura 7, PL-02-093 Warsaw, Poland}

\author{Zhoujian Cao}
\affiliation{School of Physics and Astronomy, Beijing Normal University, Beijing 100875, China}
\affiliation{School of Fundamental Physics and Mathematical Sciences, Hangzhou Institute for Advanced Study, University of Chinese Academy of Sciences, Hangzhou 310024, China}

\author{Shao-Jiang Wang}
\email{schwang@itp.ac.cn}
\affiliation{Institute of Theoretical Physics, Chinese Academy of Sciences (CAS), Beijing 100190, China}
\affiliation{Asia Pacific Center for Theoretical Physics (APCTP), Pohang 37673, Korea}

\begin{abstract}
The Hubble tension motivates direct tests of the internal consistency of the local distance ladder. We perform forward and backward consistency checks between Type Ia supernovae in Cepheid host galaxies and the Hubble-flow supernova sample using Pantheon+ data. Within flat $\Lambda$CDM, we impose Planck-CMB priors on one sample, infer the supernova absolute-magnitude calibration, and propagate it to the other sample to recover the Hubble constant, which should match the input value if the second and third rungs of the distance ladder align perfectly. We examine the impact of host-galaxy peculiar velocities using redshift corrections based on the Manticore-Local reconstruction. These corrections shift the recovered $H_0$ from $74.30^{+1.18}_{-1.20}$ to $71.37^{+1.23}_{-1.18}$ $\mathrm{km\,s^{-1}Mpc^{-1}}$ in the forward analysis and from $60.83^{+0.98}_{-0.97}$ to $63.25^{+0.95}_{-0.93}$ $\mathrm{km\,s^{-1}Mpc^{-1}}$ in the backward analysis, bringing both results closer to the input CMB constraint. Comparing the cosmological intercepts of the magnitude-redshift relation inferred from two rungs of the distance ladder before and after peculiar-velocity corrections, we find that the significance of the discrepancy decreases from $6.8\sigma$ to $3.9\sigma$ in the forward check and from $7.3\sigma$ to $4.3\sigma$ in the backward check. Peculiar-velocity corrections therefore partially alleviate the intercept tension, while the residual discrepancy is further cut in half by volumetric bias down to $\sim1.7\sigma$. Our results highlight the importance of local velocity modeling and motivate further tests of sample selection and calibration systematics in assessing distance-ladder consistency.
\end{abstract}

\maketitle

\section{Introduction}\label{sec:intro}

Recently, more evidence has pointed to the apparent contradictions between high-redshift and low-redshift observations~\cite{Perivolaropoulos:2021jda,Abdalla:2022yfr,CosmoVerseNetwork:2025alb}. The most representative cosmological tensions involve the Hubble constant $H_{0}$~\cite{Bernal:2016gxb,Verde:2019ivm,Knox:2019rjx,Riess:2019qba,DiValentino:2020zio,Freedman:2021ahq,DiValentino:2021izs,Schoneberg:2021qvd,Shah:2021onj,Verde:2023lmm,Poulin:2023lkg,Hu:2023jqc,Vagnozzi:2023nrq,Cai:2023sli,Cai:2026swf}. Among the possible methods to measure $H_{0}$, two principal methodologies are widely discussed. One employs the cosmic distance ladder, which typically utilizes geometric distance calibrations, observations of Cepheid variables, and Type Ia supernovae (SNe Ia) to build a distance measurement step by step toward $H_{0}$ measurement~\cite{Riess:2021jrx}. The other derives $H_{0}$ from measurements of the angular power spectrum of the cosmic microwave background (CMB) fluctuations~\cite{Planck:2018vyg}, which fit all free parameters globally based on the $\Lambda$-cold-dark-matter ($\Lambda$CDM) model. 

\begin{figure*}
    \centering
    \includegraphics[width=0.49\textwidth]{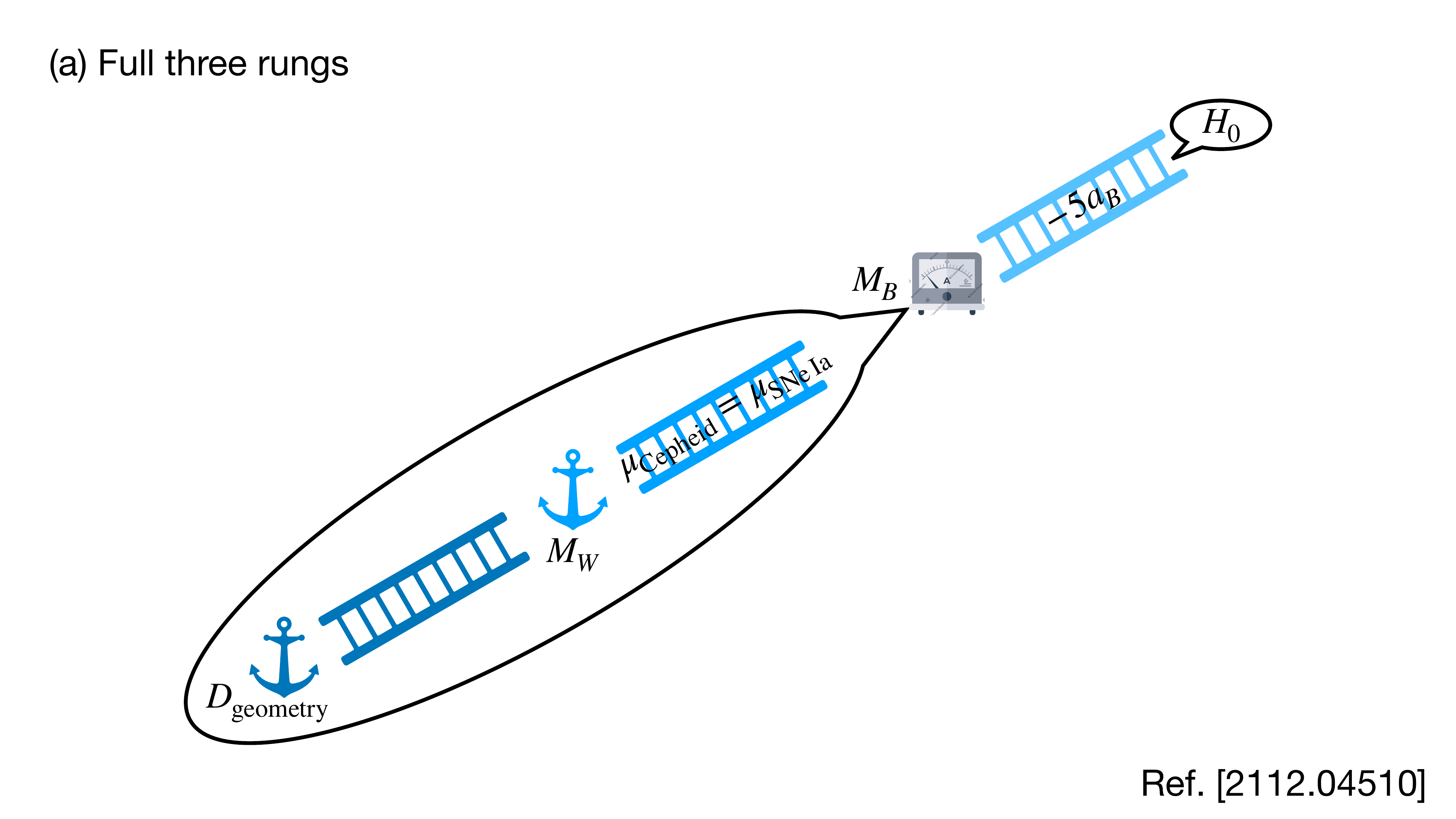}
    \includegraphics[width=0.49\textwidth]{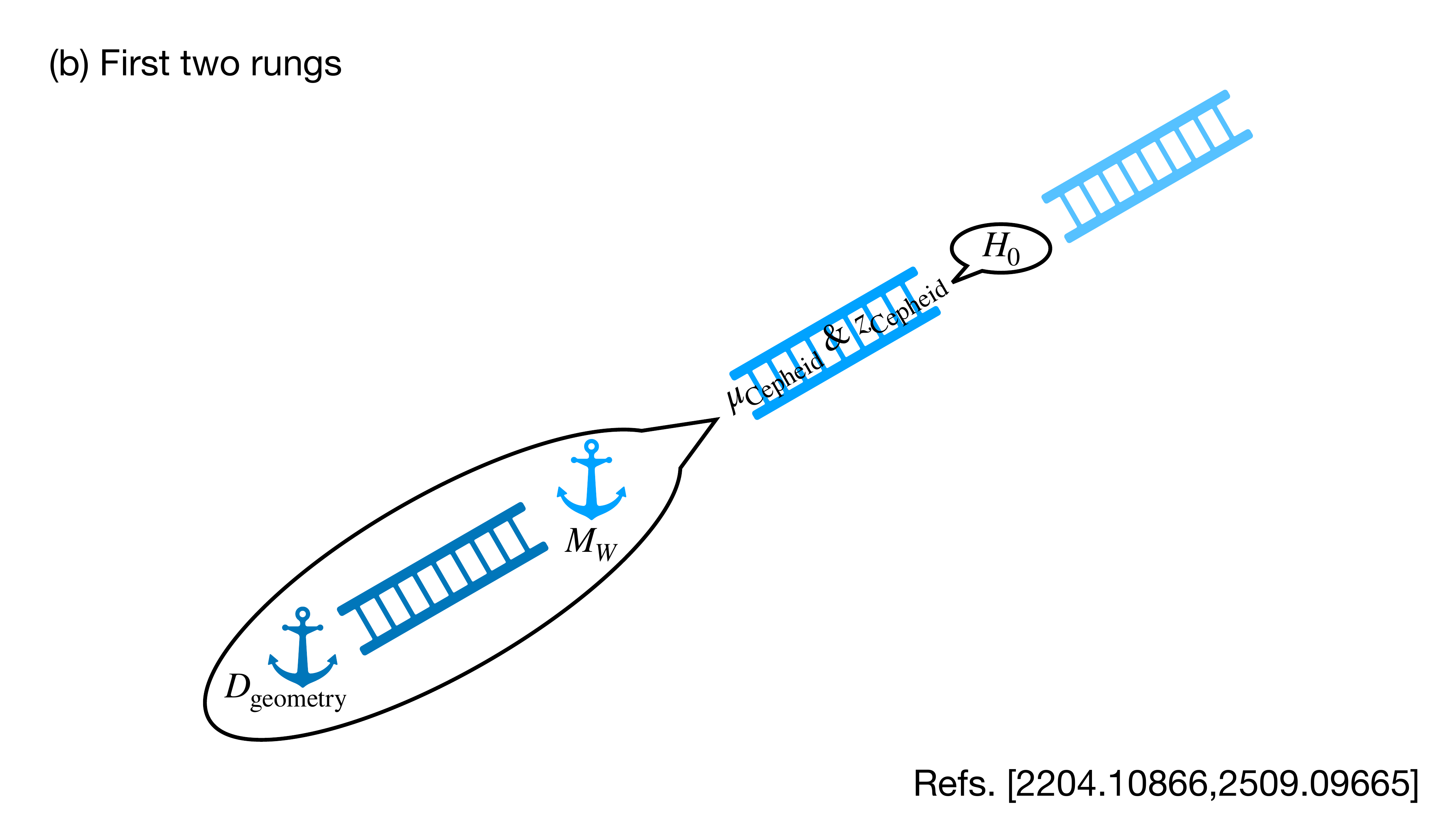}\\
    \includegraphics[width=0.49\textwidth]{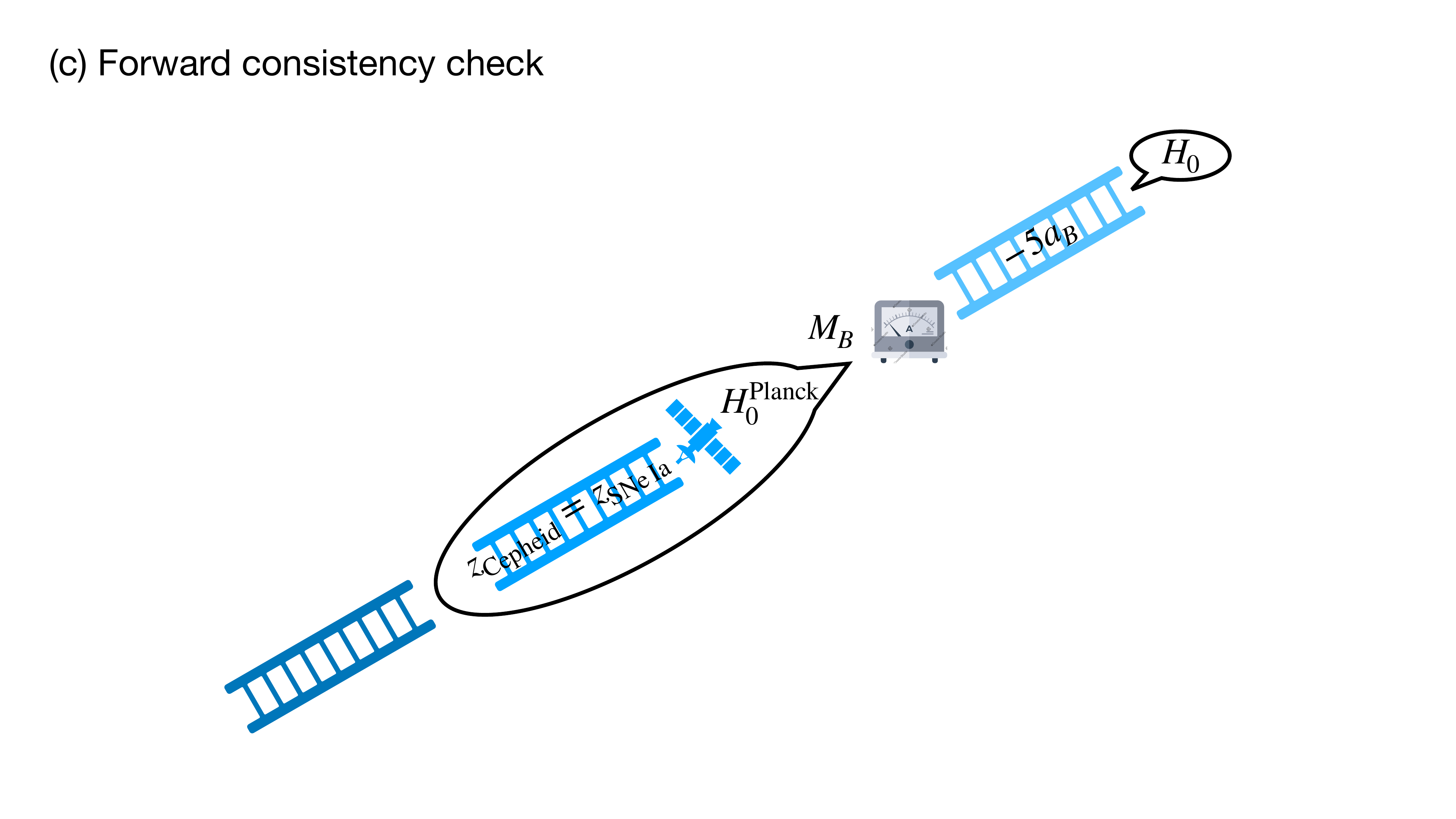}
    \includegraphics[width=0.49\textwidth]{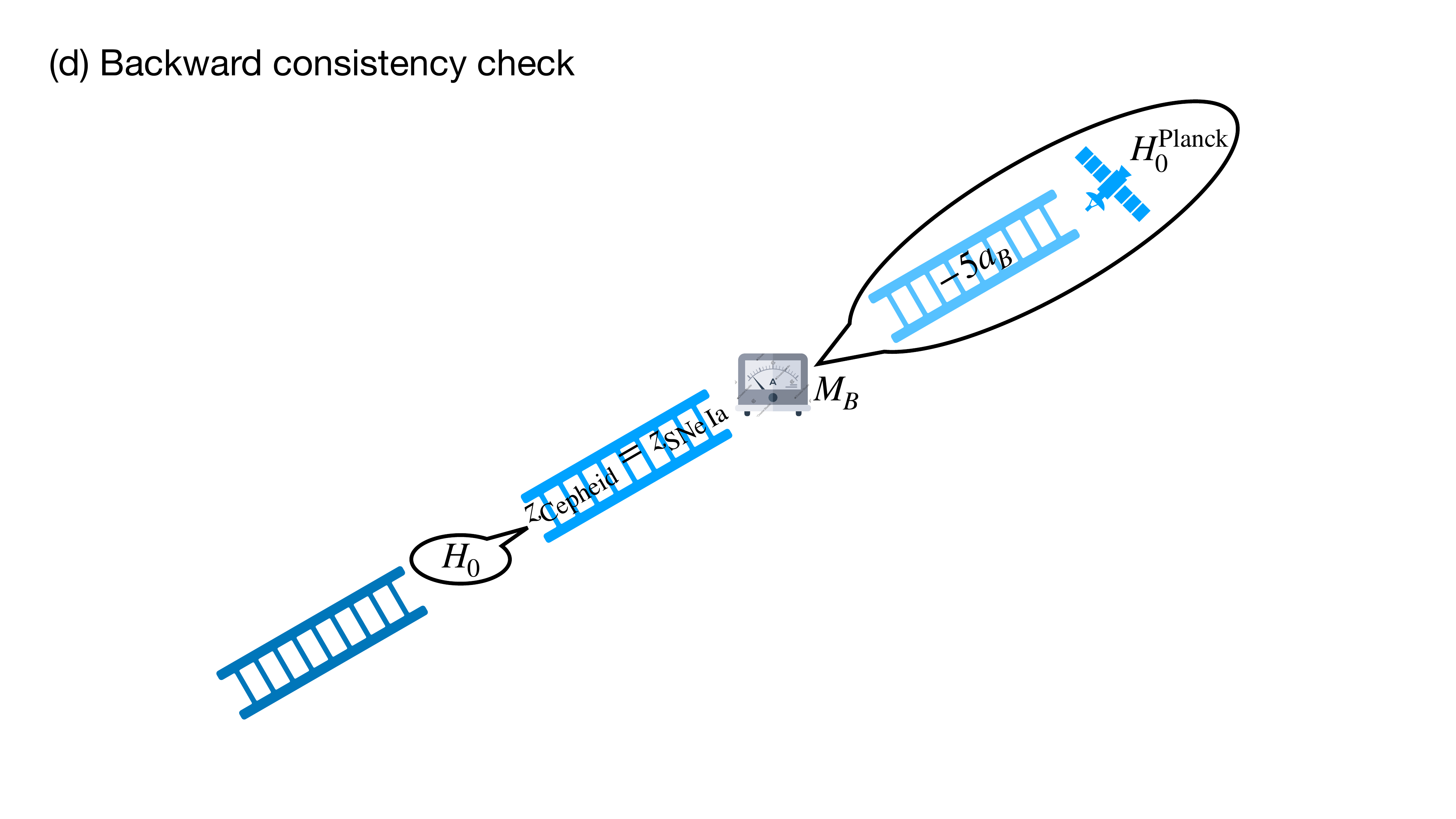}
    \caption{Schematic illustrations of the Hubble-constant measurements from (a) full-three-rung ladder and (b) first-two-rung ladder, along with the consistency checks of ladders from (c) forward and (d) backward methods as described in the main text.}
    \label{fig:consistency_check}
\end{figure*}

The cosmic distance ladder is a hierarchical method for determining distances within the local universe and comprises three distinct rungs~\cite{Riess:2022oxy}. The first rung is based on geometric methods such as parallax~\cite{Riess:2018byc,Riess:2018uxu,Riess:2020fzl} and observations of detached eclipsing binaries~\cite{Pietrzynski:2019cuz}, which provide direct distance measurements for nearby sources. The second rung employs Cepheid variable stars, whose pulsating nature leads to periodic variations in brightness, diameter, temperature, etc. These periodic variations are related to the luminosity of Cepheids~\cite{Leavitt:1912zz} by the so-called Period-Luminosity (P-L) relation, and hence one can precisely measure their distances by measuring their periods~\cite{Riess:2019cxk}. The third rung is constructed using SNe Ia as standard candles due to their high peak luminosity, which is quite uniform after proper standardization. By comparing the known absolute magnitudes of these supernovae with their observed apparent brightness, we can measure the distances to both the SNe Ia and their host galaxies accurately. Between the rungs, there are some calibrators~\cite{Scolnic:2021amr}, such as the SNe Ia located in the same galaxy as Cepheids, which serve to bridge adjacent rungs of the ladder, ensuring a coherent and stepwise determination of distances that ultimately culminates in an independent measurement of $H_0$ when combined with cosmological models. The full three-rung distance-ladder measurement of $H_0$ is illustrated in Fig.~\ref{fig:consistency_check}(a).

To test individual components of the distance ladder, an alternative approach is to determine $H_0$ using only its first two rungs~\cite{Kenworthy:2022jdh,Stiskalek:2025ktq}, as illustrated schematically in Fig.~\ref{fig:consistency_check}(b). Rather than extending the distance scale through the third-rung SNe, this approach directly compares geometrically calibrated Cepheid distances with the redshifts of their host galaxies. It therefore provides a test of the role of the supernova rung in the inferred $H_0$. There are two main results regarding this idea.

In Ref.~\cite{Kenworthy:2022jdh}, they implemented this approach using 35 SH0ES Cepheid hosts with $cz\lesssim3300\,\mathrm{km\ s^{-1}}$. They combined Cepheid distance moduli calibrated geometrically in the Milky Way, NGC4258, and the Large Magellanic Cloud and their covariance with host-galaxy redshifts in a hierarchical Bayesian model. Since they select low-redshift samples, peculiar velocities are a significant source of systematic uncertainty. They present a statistical formalism for mitigating and quantifying their effects. Besides, they consider both distance-limited and redshift-limited sample selection. An equally weighted combination of four preferred models yielded $H_0=73.1^{+2.6}_{-2.3}\,\mathrm{km\ s^{-1}\ Mpc^{-1}}$, consistent with the three-rung SH0ES determination and in $2.6\sigma$ tension with Planck.

In Ref.~\cite{Stiskalek:2025ktq}, they subsequently revisited the same 35-host sample using a Bayesian forward model of the Cepheid magnitudes, periods and metallicity, geometric-anchor constraints, and host-galaxy redshifts. Their analysis directly modeled the Cepheid observables, introduced an explicit treatment of sample selection, and employed the \textit{Bayesian Origin Reconstruction from Galaxies based}(BORG-based) \texttt{Manticore-Local} density and velocity reconstruction to account for local structure and its uncertainty. Applying the selection by SN apparent magnitude, their updated analysis yielded $H_0=71.1\pm1.4\,\mathrm{km\ s^{-1}\ Mpc^{-1}}$, achieving approximately 2 percent statistical precision. Although slightly lower than that of Ref.~\cite{Kenworthy:2022jdh}, this result still remains in $2.8\sigma$ tension with the Planck-CMB $H_0$ constraint.

Alongside these two-rung determinations of $H_0$, recent studies have found a dubbed cosmological intercept tension and its implications for cosmological inference~\cite{Huang:2024erq,Huang:2024gfw,Huang:2025som,Ling:2025lmw,Wang:2026kor}. These discussions motivated us to make a direct examination of the consistency between the second and third rungs of the distance ladder. In this work, we perform both forward and backward consistency checks by inputting the CMB prior to propagate the $M_B$ calibration between the two rungs and examining the resulting $H_0$ constraints and intercept posteriors, as illustrated schematically in Fig.~\ref{fig:consistency_check}(c) and Fig.~\ref{fig:consistency_check}(d). Besides, we also consider the influence of redshift corrections from peculiar-velocity reconstructions for calibrators mentioned in Ref.~\cite{Stiskalek:2025ktq}.

The outline is as follows. We first describe our forward/backward methodology in Sec.~\ref{sec:methodology}, then show the results in Sec.~\ref{sec:data}, and check for volumetric bias in Sec.~\ref{app:bias}. Finally, we conclude with some discussion in Sec.~\ref{sec:conclusion}.

\section{Distance ladder consistency} \label{sec:methodology}

We compare two Pantheon+ samples: the calibrator SNe Ia in Cepheid host galaxies, denoted $\mathcal{C}$, and the Hubble-flow SNe Ia, denoted $\mathcal{F}$. The parent calibrator sample contains 42 SNe Ia~\cite{Scolnic:2021amr}; the subset with external host-velocity estimates is described in Sec.~\ref{subsec:redshift_corrections}. Although these objects conventionally connect the second and third rungs, we infer $M_B$ here from their magnitudes and redshifts under the Planck-CMB prior on $H_0$, rather than from a fit to Cepheid distance moduli. Both checks use the following distance model and likelihood.

In flat $\Lambda$CDM, the luminosity distance is
\begin{equation}
\label{1}
D_{\rm L}(z_{\rm HD},z_{\rm hel})=(1+z_{\rm hel})\int_0^{z_{\rm HD}}\frac{c\,\mathrm{d}z'}{H(z')},
\end{equation}
with the Hubble function
\begin{equation}
\label{2}
H(z)=H_0\sqrt{\Omega_{\rm m}(1+z)^3+1-\Omega_{\rm m}}.
\end{equation}
Here, $c$ is the speed of light, $z_{\rm hel}$ is the heliocentric redshift that is observed in the heliocentric frame, and $z_{\rm HD}$ is the Hubble-diagram redshift that is corrected for the solar motion relative to the CMB rest frame and the peculiar velocity of the supernova host galaxy, which is used to describe the cosmological expansion. We follow the notation of \cite{Kenworthy:2019qwq}, where they note that the $1+z$ factor in the luminosity distance definition arises entirely from the loss of photon energy due to redshift and should therefore be calculated using the apparent redshift rather than a redshift corrected to the cosmological rest frame. Consequently, the heliocentric redshift enters the prefactor, while the Hubble-diagram redshift sets the integration limit (see Eq.(16) of Ref.~\cite{Kenworthy:2019qwq}). 


We use the standardized magnitudes $m_B^{\rm corr}$ released in Ref.~\cite{Brout:2022vxf}, which incorporate light-curve, color, selection-bias, and host-mass corrections~\cite{SNLS:2010pgl,Brout:2021mpj,Brout:2022vxf}. Thus, the observed distance modulus is
\begin{equation}
\label{3}
\mu_i=m_{B,i}^{\rm corr}-M_B,
\end{equation}
and its model prediction is
\begin{equation}
\label{7}
\mu_{{\rm model},i}=5\log_{10}\!\left[\frac{D_{\rm L}(z_{{\rm HD},i},z_{{\rm hel},i})}{10\,\mathrm{pc}}\right].
\end{equation}
We do not refit the light-curve standardization parameters. For each sample, the supernova likelihood is~\cite{SupernovaSearchTeam:1998fmf,SNLS:2005qlf,SNLS:2011lii}
\begin{equation}
\label{5}
-2\ln\mathcal{L}_{\rm SN}=\Delta\boldsymbol{\mu}^{\mathsf T}C_{\rm stat+syst}^{-1}\Delta\boldsymbol{\mu}+\mathrm{const.},
\end{equation}
where
\begin{equation}
\label{6}
\Delta\mu_i=\mu_i-\mu_{{\rm model},i},
\end{equation}
and $C_{\rm stat+syst}$ is the covariance matrix taken from the Pantheon+ release for the selected data~\cite{Scolnic:2021amr,Brout:2022vxf}.We infer the joint posterior of $(M_B,H_0,\Omega_{\rm m})$ by combining this likelihood with the priors appropriate to each stage of fitting. For the first stage, the priors are the Planck-CMB priors on $H_0 = \mathcal{N}(67.36,0.54)\,\mathrm{km\ s^{-1}\ Mpc^{-1}}$ and $\Omega_{\rm m} = \mathcal{N}(0.3153,0.0073)$~\cite{Planck:2018vyg}. For the second stage, they are marginalized over to get the posterior of $M_B$.

As a diagnostic to utilize the $M_B$--$H_0$ degeneracy, we also evaluate the cosmological intercept $a_B$ defined by isolating the dimensionless luminosity distance $d_L(z)=D_L(z)/(c/H_0)$ via $m_{B,i}\equiv 5\lg d_L(z_i)-5a_B$ as
\begin{equation}
\label{13}
-5a_B=M_B+5\log_{10}\!\left[\frac{c/H_0}{\mathrm{Mpc}}\right]+25.
\end{equation}
The intercept posteriors are obtained from the joint posterior samples at each stage of fitting, retaining the correlation between $M_B$ and $H_0$. Since $-5a_B$ depends on the same parameter combination that controls the supernova magnitude--redshift relation, it provides a complementary representation of the mismatch rather than an independent measurement.

\subsection{Forward consistency check}\label{subsec:forward}

The forward check transfers the calibration from $\mathcal{C}$ to $\mathcal{F}$ [Fig.~\ref{fig:consistency_check}(c)]. In the first stage, we fit the calibrator sample $\mathcal{C}$ using the likelihood in Eq.~\eqref{5} and Planck priors on $H_0 = \mathcal{N}(67.36,0.54)\,\mathrm{km\ s^{-1}\ Mpc^{-1}}$ and $\Omega_{\rm m} = \mathcal{N}(0.3153,0.0073)$~\cite{Planck:2018vyg}. Marginalizing the joint posterior over these two parameters yields the calibration distribution $p_{\mathcal{C}}(M_B)$, as we report later in the results. 
In the second stage, we fit the Hubble-flow sample $\mathcal{F}$ using $p_{\mathcal{C}}(M_B)$ as the calibration prior. Marginalizing over $M_B$ and $\Omega_{\rm m}$ gives the recovered $H_0$ posterior, which we compare with the input CMB constraint. We also compare the intercept posteriors inferred at these two stages.

\subsection{Backward consistency check}\label{subsec:backward}

The backward check transfers the calibration from $\mathcal{F}$ to $\mathcal{C}$ [Fig.~\ref{fig:consistency_check}(d)]. In the first stage, we fit the Hubble-flow sample $\mathcal{F}$ using the likelihood in Eq.~\eqref{5} and the same Planck priors on $H_0$ and $\Omega_{\rm m}$. Marginalizing the joint posterior over these two parameters yields the calibration distribution $p_{\mathcal{F}}(M_B)$, as we report later in the results.
In the second stage, we fit the calibrator sample $\mathcal{C}$ using $p_{\mathcal{F}}(M_B)$ as the calibration prior. Marginalizing over $M_B$ and $\Omega_{\rm m}$ gives the recovered $H_0$ posterior. As in the forward check, we compare it with the input CMB constraint and compare the intercept posteriors at these two stages.

\subsection{Peculiar velocity corrections}\label{subsec:redshift_corrections}

Most calibrators lie at $cz\lesssim3000\,\mathrm{km\,s^{-1}}$, where peculiar velocities can contribute significantly to the observed redshift. We assess this sensitivity using the CMB-frame host redshifts and line-of-sight peculiar velocities tabulated in Table D1 of Ref.~\cite{Stiskalek:2025ktq}. For nonrelativistic peculiar velocities, the corrected redshift is
\begin{equation}
\label{8}
z_{\rm HD}=\frac{1+z_{\rm CMB}}{1+V_{\rm pec}/c}-1,
\end{equation}
where $z_\mathrm{CMB}$ is CMB corrected redshift, and the positive $V_{\rm pec}$ denotes motion away from the observer in CMB-frame. 
We adopt the tabulated \texttt{Manticore-Local} velocity estimates and substitute the resulting $z_{\rm HD}$ into Eq.~\eqref{1}.
The velocity table covers 35 Cepheid host galaxies. Host corrections are therefore assigned to their associated SNe, and the corrected analysis is restricted to this matched subset of the parent calibrator sample. We compare the reference redshift treatment, labeled ``(w/o) without correction'' below, with this corrected treatment described above, labeled ``(w/) with correction.'' This comparison assesses the effect of the adopted host corrections; uncertainties in the velocity estimates and sample matching must also be considered when interpreting the shift of the intercept $-5a_{B}$ between the two rungs.

\section{Results of consistency checks} \label{sec:data}

\begin{figure}[t]
    \centering
    \includegraphics[width=\linewidth]{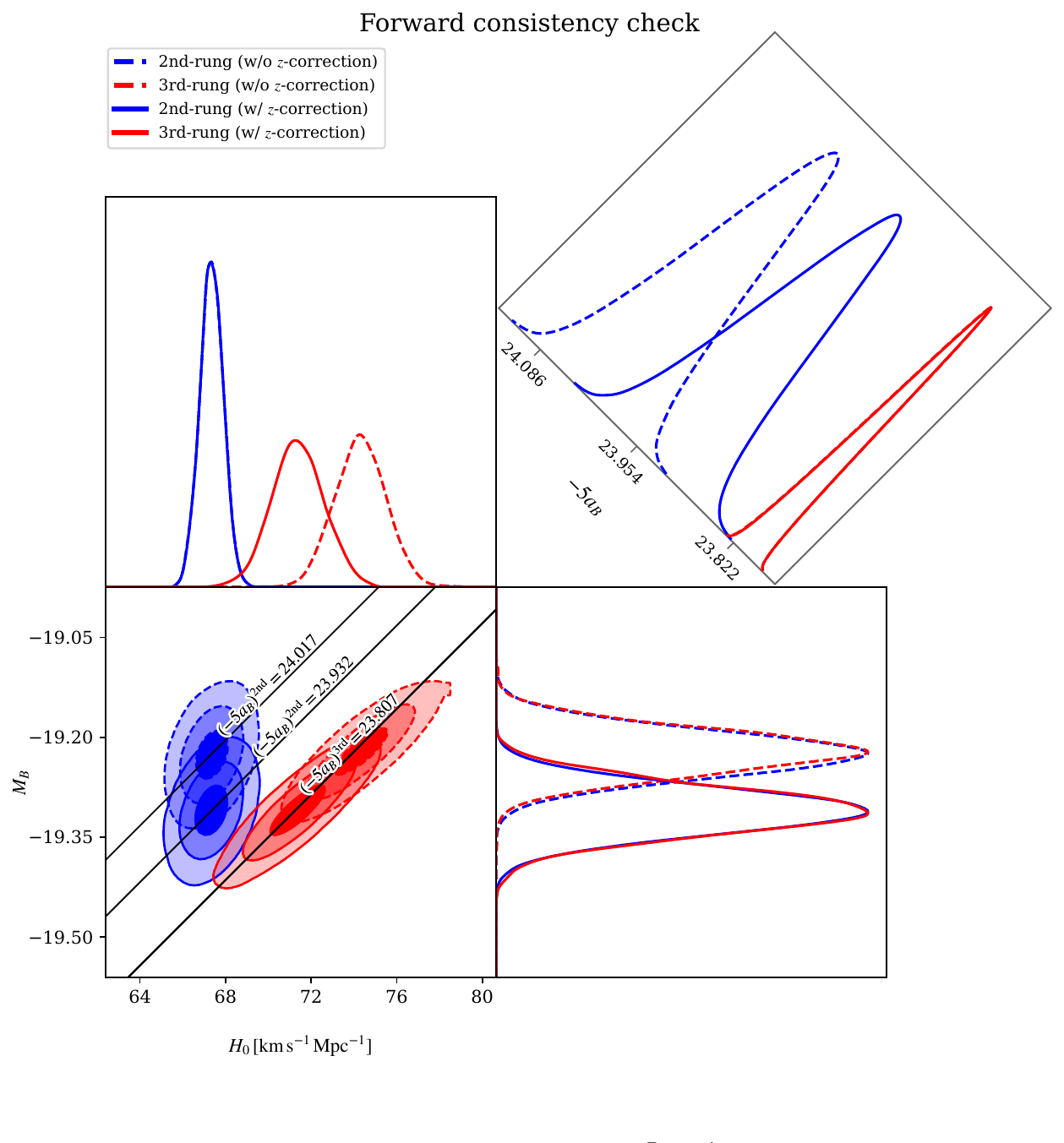}
    \caption{Forward consistency check from input/output $H_0$ constraints at second/third rungs (blue/red) with intermediate $M_B$ constraints, which combine into the intercept ($-5a_B$) constraints in the tilted inset, before/after the peculiar-velocity corrections (dashed/solid).}
    \label{fig:forward_result}
\end{figure}
\begin{figure}[t]
    \centering
    \includegraphics[width=\linewidth]{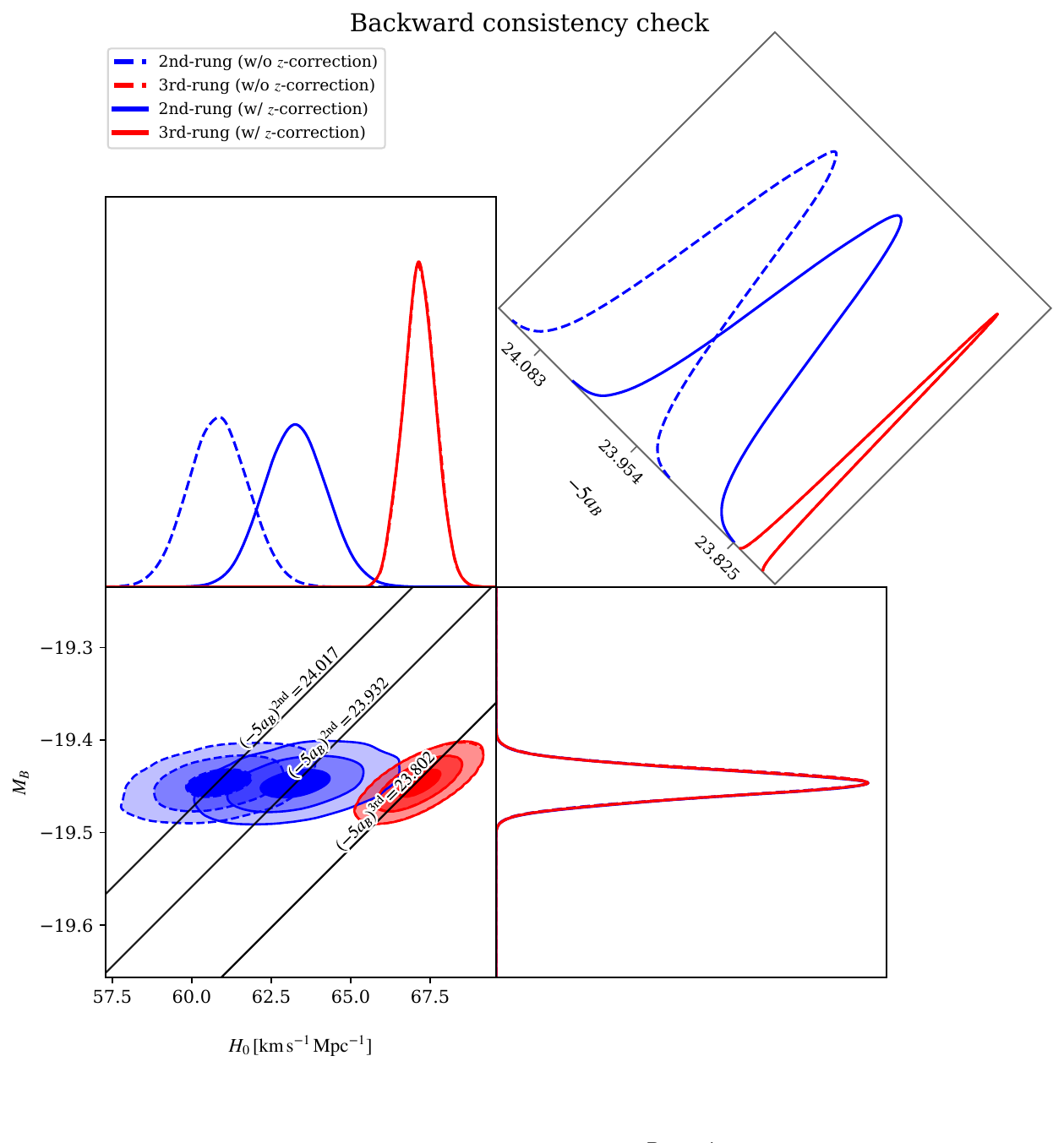}
    \caption{Backward consistency check from input/output $H_0$ constraints at third/second rungs (red/blue) with intermediate $M_B$ constraints, which combine into the intercept ($-5a_B$) constraints in the tilted inset, before/after the peculiar-velocity corrections (dashed/solid).}
    \label{fig:backward_result}
\end{figure}

\subsection{Forward consistency check} \label{subsec:forward_result}

We present the results of the forward and backward consistency checks separately, beginning with the forward analysis. We first examine the marginalized $M_B$ posterior inferred from the calibrator sample $\mathcal{C}$, which provides the calibration prior for the subsequent determination of $H_0$
from the Hubble-flow SNe Ia sample $\mathcal{F}$.

Without the redshift corrections (w/o $z$-correction), we obtain
\begin{equation}
\label{9}
    M_B=-19.225\pm0.035\;\mathrm{mag}.
\end{equation}
With the redshift corrections (w/ $z$-correction), the constraint becomes
\begin{equation}
\label{10}
    M_B=-19.310\pm0.035\;\mathrm{mag}.
\end{equation}
Applying the redshift correction shifts the $M_B$ posterior towards more negative values, corresponding to a brighter absolute-magnitude calibration. This shift reflects the sensitivity of the calibrator-based calibration to the treatment of peculiar velocities at low redshifts. When propagated to the Hubble-flow sample, the brighter calibration leads to a lower inferred $H_0$.

Marginalizing the second-stage joint posterior over $M_B$ and $\Omega_m$,
we obtain
\begin{equation}
\label{11}
    H_0=74.30^{+1.18}_{-1.20}\;\mathrm{km\ s^{-1}\ Mpc^{-1}}
\end{equation}
without the redshift corrections, and
\begin{equation}
\label{12}
    H_0=71.37^{+1.23}_{-1.18}\;\mathrm{km\ s^{-1}\ Mpc^{-1}}
\end{equation}
with the corrections.

Although the redshift correction brings the recovered $H_0$ posterior closer to the CMB-inferred constraint used in the initial calibration, the results obtained with and without the redshift corrections both remain inconsistent with the $H_0$ value measured from CMB analysis. This residual $H_0$ discrepancy suggests a potential mismatch between the second and third rungs of the distance ladder.

Besides, motivated by the $a_B$ tension between the second and third rungs reported in Refs.~\cite{Huang:2024erq,Huang:2024gfw} for the local and late-time Universe, we further examine the cosmological intercept $-5a_B$ as a complementary diagnostic of the discrepancy seen in our $H_0$ and $M_B$ constraints. Specifically, we derive the intercept distributions from the joint posterior samples obtained at each of these two stages of the forward analysis, and then compare two results (w/ correction and w/o redshift corrections) for each rung separately. For the second rung, the results are
\begin{equation}
\label{14}
\begin{aligned}
    -5a_B&=24.017 \pm 0.030\quad\text{(w/o $z$-corrections)},\\
    -5a_B&=23.932 \pm 0.030\quad\text{(w/ $z$-corrections)}.
\end{aligned}
\end{equation}
For the third rung, the results are
\begin{equation}
\label{15}
\begin{aligned}
    -5a_B&=23.807 \pm 0.007\quad\text{(w/o $z$-corrections)},\\
    -5a_B&=23.807 \pm 0.007\quad\text{(w/ $z$-corrections)}.
\end{aligned}
\end{equation}
These results are summarized in a tilted inset of Fig.~\ref{fig:forward_result}. Applying the host-galaxy redshift correction for peculiar velocity brings the second-rung intercept closer to the third-rung intercept, thus reducing the significance of the discrepancy between the second- and third-rung $-5a_B$ constraints from $6.8\sigma$ to $3.9\sigma$. This reduction shows that the inferred level of agreement between the two rungs is sensitive to the treatment of host-galaxy redshifts. Although the correction partially alleviates the intercept tension, a residual discrepancy remains, suggesting a potential mismatch between these two rungs within our current analysis framework.

\subsection{Backward consistency checks} \label{subsec:backward_result}

We now turn to the results of the backward consistency check, in which the $M_B$ calibration inferred from the third-rung Hubble-flow SNe Ia sample $\mathcal{F}$ is used to constrain $H_0$ with the second-rung calibrator sample $\mathcal{C}$. We briefly summarize the resulting $M_B$ and $H_0$ constraints, followed by a comparison of the $-5a_B$ posteriors derived from these two stages.

For $M_B$, we obtain
\begin{equation}
\label{16}
    M_B=-19.447\pm0.014\;\mathrm{mag}.
\end{equation}
from the third-rung sample, regardless of the redshift corrections that are applied to the second-rung sample. But for $H_0$, the constraints will be different. We obtain
\begin{equation}
\label{17}
    H_0=60.83^{+0.98}_{-0.97}\;\mathrm{km\ s^{-1}\ Mpc^{-1}}
\end{equation}
without the redshift corrections, and
\begin{equation}
\label{18}
    H_0=63.25^{+0.95}_{-0.93}\;\mathrm{km\ s^{-1}\ Mpc^{-1}}
\end{equation}
with the redshift corrections.
Combining together the forward and backward analyses shows that applying the redshift corrections brings the recovered $H_0$ constraints closer to the CMB-inferred value. Nevertheless, substantial discrepancies remain under both redshift treatments, indicating an inconsistency between the second and third rungs of the distance ladder.

We next examine whether this discrepancy is also reflected in the cosmological intercept $-5a_B$. As in the forward analysis, we compare the intercept posteriors derived from the two stages of the backward consistency check. For the second rung, the results are
\begin{equation}
\label{19}
\begin{aligned}
    -5a_B&=24.017 \pm 0.030\quad\text{(w/o $z$-corrections)},\\
    -5a_B&=23.932 \pm 0.030\quad\text{(w/ $z$-corrections)}.
\end{aligned}
\end{equation}
For the third rung, the results are
\begin{equation}
\label{20}
\begin{aligned}
    -5a_B&=23.802 \pm 0.004\quad\text{(w/o $z$-corrections)},\\
    -5a_B&=23.802 \pm 0.004\quad\text{(w/ $z$-corrections)}.
\end{aligned}
\end{equation}
These results are also summarized in a tilted inset of Fig.~\ref{fig:backward_result}. Applying the host-galaxy redshift correction brings the second-rung intercept closer to the third-rung intercept, thus reducing the significance of the discrepancy between the second- and third-rung $-5a_B$ constraints from $7.3\sigma$ to $4.3\sigma$. These results are similar to those of the forward consistency check and likewise indicate a potential mismatch between the second and third rungs of the distance ladder.

\section{Volumetric bias}\label{app:bias}

The above analysis applies Manticore-Local peculiar-velocity estimates to the Cepheid-host redshifts, but it does not implement a hierarchical distance prior or explicitly model the selection of the nearby SN hosts by volume-weighted scattering in redshift, a form of Malmquist bias, as done in Refs.~\cite{Brout:2022vxf,Kenworthy:2022jdh}, though roughly assuming an uncorrected peculiar-velocity dispersion of $\sigma_v=250\,\mathrm{km\,s^{-1}}$. The Pantheon+ covariance matrix does not automatically correct for the resulting conditional-distance bias. Thus, a remaining question is whether this Malmquist-type effect can explain approximately 0.13 mag, corresponding to roughly a $6\%$ distance difference. We will carry out this essential check below.

\subsection{Analytic estimations}

The physical mechanism behind this Malmquist-type bias is that, at fixed observed redshift, there can be more distant galaxies scattered toward lower redshifts than nearer galaxies scattered toward higher redshifts, because the available volume increases quadratically with radius. Therefore, even though positive and negative peculiar velocities are equally probable, the number of galaxies available to scatter downward from a larger cosmological redshift is greater than the number available to scatter upward from a smaller one. Hence, there is a bias.

Specifically, to model the observed redshift $z_\mathrm{obs}$ in the CMB frame from the true cosmological redshift $z$ with the line-of-sight peculiar velocity $v$ of distribution $v\sim\mathcal{N}(0,\sigma_v^2)$, one has $cz_\mathrm{obs}\simeq cz+v$ at low redshift. 
Then, the likelihood for the observed redshift $z_\mathrm{obs}$ given the true redshift $z$ is
\begin{align}
P(z_\mathrm{obs}\mid z)=\frac{1}{\sqrt{2\pi}\sigma_v}\exp\left[-\frac{c^2(z_\mathrm{obs}-z)^2}{2\sigma_v^2}\right].
\end{align}
For a homogeneous population with constant comoving number density, the distance prior is $P(r)\mathrm{d}r\propto r^2\mathrm{d}r$, or equivalently $P(z)\propto z^2$ at low redshift. Bayes' theorem then gives
\begin{align}
&P(z\mid z_\mathrm{obs})=\frac{P(z_\mathrm{obs}\mid z)P(z)}{\displaystyle\int_0^\infty P(z_\mathrm{obs}\mid z')P(z')\mathrm{d}z'},\\
&=\frac{z^2\exp[-(z-z_\mathrm{obs})^2/(2\sigma_z^2)]}
{\displaystyle\int_0^\infty
z'^2\exp[-(z'-z_\mathrm{obs})^2/(2\sigma_z^2)]\,\mathrm{d}z'},\quad \sigma_z\equiv\frac{\sigma_v}{c}.\nonumber
\end{align}
Thus, the Gaussian likelihood is symmetric about $z=z_\mathrm{obs}$, but multiplication by the volume factor $z^2$ shifts the posterior $P(z\mid z_{\rm obs})$ toward larger true distances, $\langle z|z_{\rm obs}\rangle>(cz_{\rm obs})/H_0$, and hence, the supernova therefore tends to be fainter than expected from the naive redshift distance, producing a positive Hubble residual, $\Delta\mu=\mu_{\rm SN}-\mu_{\rm model}(z_{\rm obs})>0$. This has the same sign as the reported nearby-sample offset: a larger magnitude intercept than for the Hubble-flow sample.

To explicitly derive the redshift dependence analytically, we first calculate the mean true redshift at fixed observed redshift:
\begin{align}
\langle z\mid z_\mathrm{obs}\rangle=
\frac{\displaystyle\int_0^\infty z^3e^{-(z-z_\mathrm{obs})^2/(2\sigma_z^2)}dz}
{\displaystyle\int_0^\infty z^2e^{-(z-z_\mathrm{obs})^2/(2\sigma_z^2)}dz}.
\end{align}
For $\sigma_z\ll z_\mathrm{obs}$ with expansion around $z=z_\mathrm{obs}$, extending the integration limits to $(-\infty,\infty)$ gives
\begin{align}
\langle z\mid z_\mathrm{obs}\rangle\simeq z_\mathrm{obs}\frac{z_\mathrm{obs}^2+3\sigma_z^2}{z_\mathrm{obs}^2+\sigma_z^2}
\simeq z_\mathrm{obs}\left(1+2\frac{\sigma_z^2}{z_\mathrm{obs}^2}\right).
\end{align}
Thus the bias in the mean true redshift is approximately
\begin{align}
\frac{\langle z\rangle-z_{\rm obs}}{z_{\rm obs}}\simeq 2\left(\frac{\sigma_v}{cz_{\rm obs}}\right)^2.
\end{align}

When measured in magnitudes or the distance modulus $\mu(z)=5\lg[D_L(z)/10\,\mathrm{pc}]$ with $D_L(z)\simeq cz/H_0$ at low redshift, the mean logarithmic distance $\langle\Delta\mu\rangle\equiv\langle\mu(z)-\mu(z_\mathrm{obs})\rangle=\langle5\lg(z/z_\mathrm{obs})\rangle_{z\mid z_\mathrm{obs}}$ is calculated by
\begin{align}\label{eq:app-bias}
\langle\Delta\mu\rangle=\frac{5}{\ln10}
\frac{\displaystyle\int_0^\infty z^2e^{-(z-z_{\rm obs})^2/(2\sigma_z^2)}\ln\frac{z}{z_{\rm obs}}\,\mathrm{d}z}
{\displaystyle\int_0^\infty z^2e^{-(z-z_{\rm obs})^2/(2\sigma_z^2)}\mathrm{d}z}.
\end{align}
This is the exact integral within the homogeneous, Gaussian, low-redshift model. It can be evaluated numerically without a series expansion. For small $\sigma_z/z_{\rm obs}$, one obtains an order-of-magnitude estimate of this bias as
\begin{align}
b_\mu(z_\mathrm{obs})\equiv\langle\Delta\mu\rangle\simeq\frac{15}{2\ln10}\left(\frac{\sigma_v}{cz_\mathrm{obs}}\right)^2.
\end{align}

\subsection{Numerical evaluations}

\begin{figure}
    \centering
    \includegraphics[width=\linewidth]{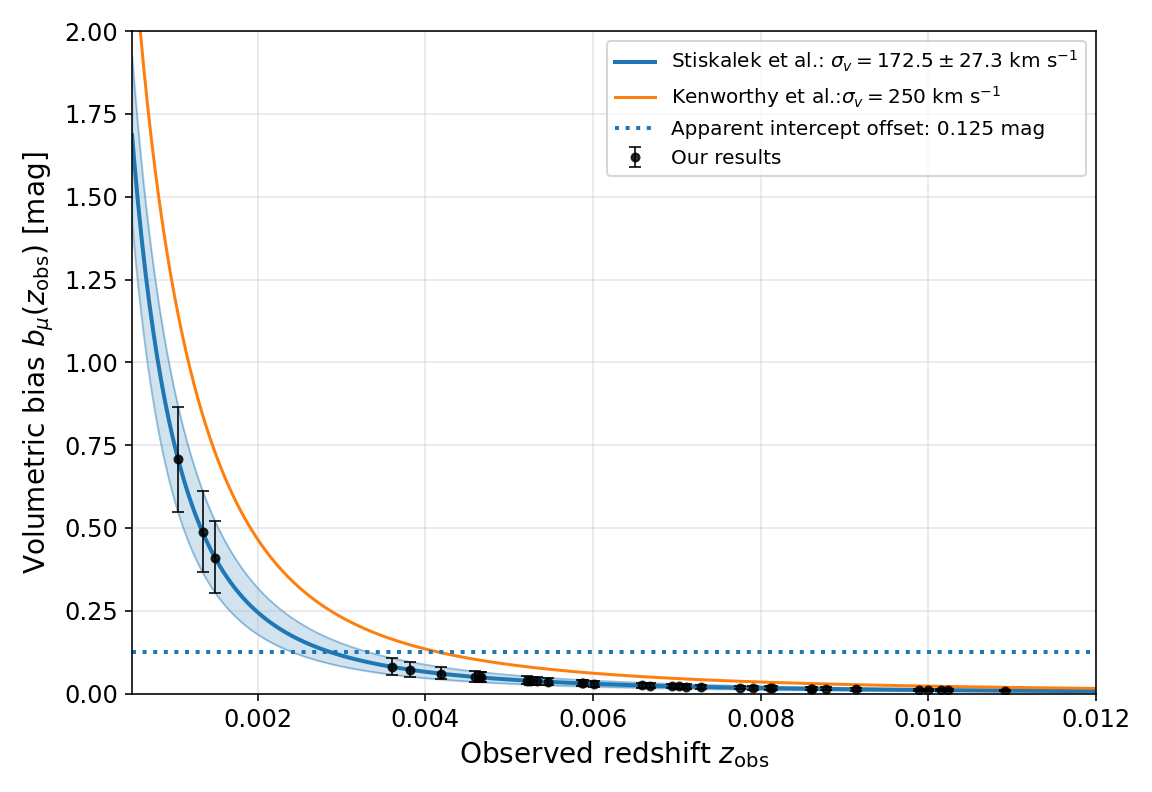}
    \caption{The volumetric bias estimates with respect to the observed redshift for the residual peculiar-velocity dispersions from Kenworthy et al. ($\sigma_v=250\,\mathrm{km\,s^{-1}\,Mpc^{-1}}$) and Stiskalek et al. ($\sigma_v=172.5\pm27.3\,\mathrm{km\,s^{-1}\,Mpc^{-1}}$), compared to the apparent intercept offset $\sim0.125$ mag. reported above.}
    \label{fig:volumetricbias}
\end{figure}

\begin{figure*}
    \centering
    \includegraphics[width=0.49\linewidth]{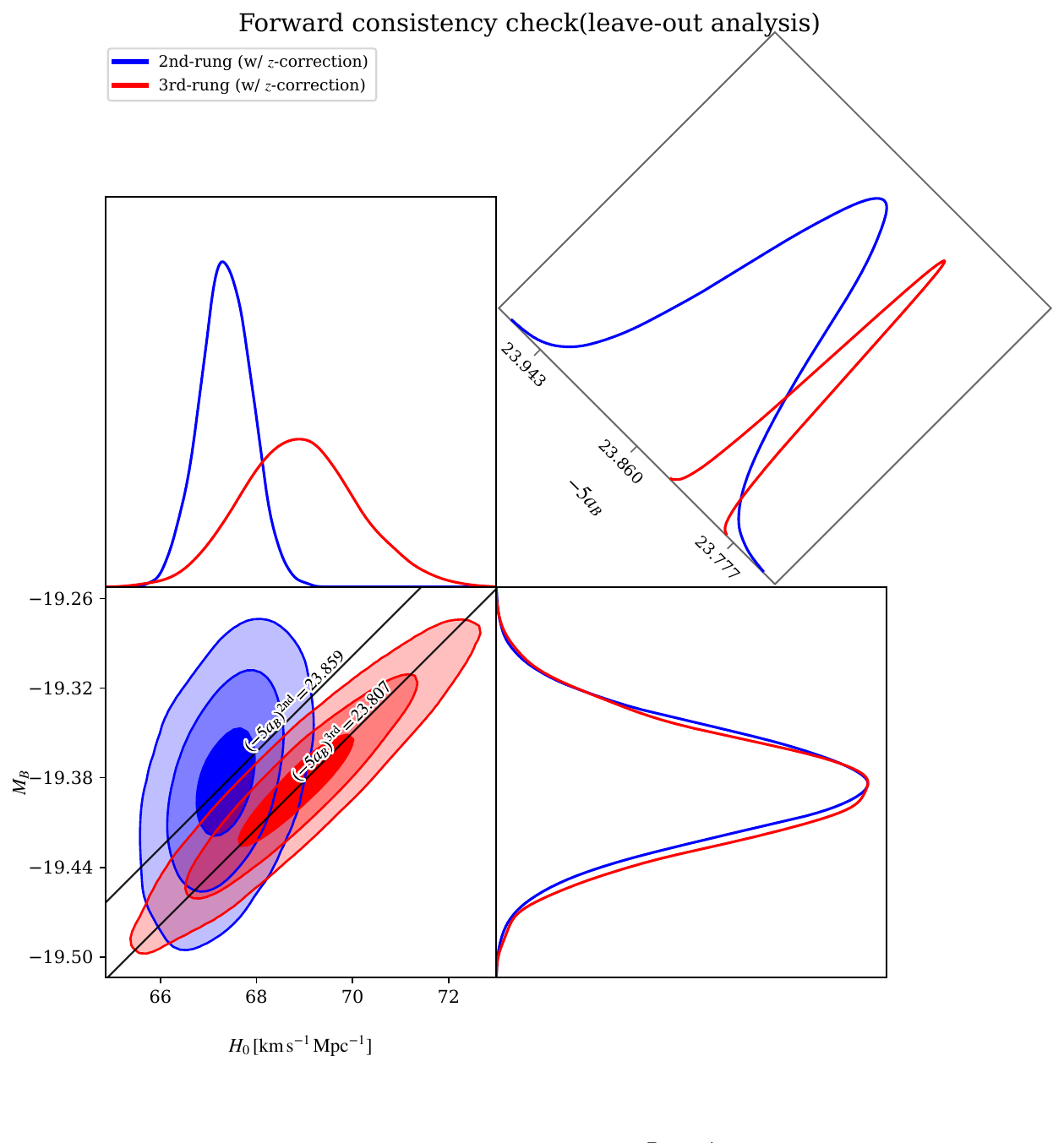}
    \includegraphics[width=0.49\linewidth]{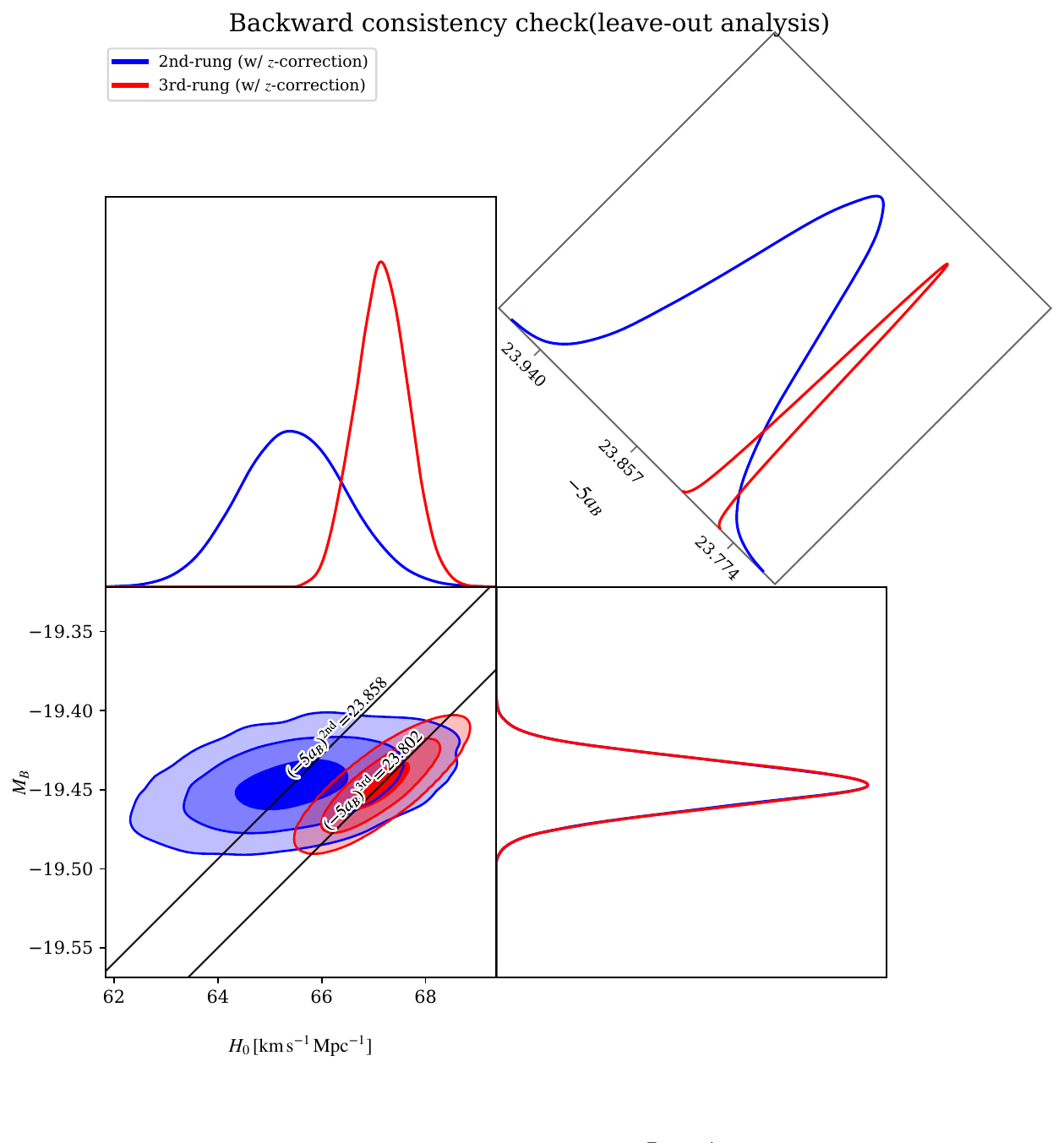}\\
    \caption{The same forward and backward consistency checks as Fig.~\ref{fig:forward_result} and Fig.~\ref{fig:backward_result} in the leave-out analysis.}
    \label{fig:leaveoutcheck}
\end{figure*}

For a typical $\sigma_v=250\,\mathrm{km\,s^{-1}}$~\cite{Kenworthy:2022jdh}, we arrive at the familiar form of this volumetric bias for the modulus,
\begin{align}
b_\mu(z_\mathrm{obs})\simeq\frac{2.265\times10^{-6}}{z_\mathrm{obs}^2}\,\mathrm{mag}.,
\end{align}
which roughly reproduces a previous estimate,
\begin{align}
b_\mu(z_\mathrm{obs})\simeq
\frac{5}{\ln10}\frac{\Delta z_\mathrm{obs}}{z_\mathrm{obs}}
=\frac{2.1715\times10^{-6}}{z^2}\ \mathrm{mag}.,
\end{align}
from Kenworthy \textit{et al.}'s~\cite{Kenworthy:2022jdh} Monte Carlo approximation $\Delta z_\mathrm{obs}/z_\mathrm{obs}\sim10^{-6}/z_\mathrm{obs}^2$. This offset in distance modulus can be transferred to the offset in the intercept as
\begin{align}
\Delta\mu&=\Delta\left(m-M-5\lg\frac{D_L(z)}{10\,\mathrm{pc}}\right)\nonumber\\
&=\Delta\left(\left[m-5\lg d_L(z)\right]-\left[M+5\lg\frac{c/H_0}{10\,\mathrm{pc}}\right]\right)\nonumber\\
&=\Delta(-5a_B).
\end{align}

As a preliminary test, we can calculate the leading uncertainty to the fitted intercept as
\begin{align}
\Delta(-5a_B)=\frac{\boldsymbol 1^TC^{-1}\boldsymbol b}{\boldsymbol 1^TC^{-1}\boldsymbol 1}
\end{align}
from the volumetric bias vector $b_i=b_\mu(z_{\mathrm{HD},i};\sigma_{v,\rm residual})$ using Eq.~\eqref{eq:app-bias} as shown in Fig.~\ref{fig:volumetricbias}, where $C$ is the Pantheon+ statistical and systematic covariance matrix, and $z_{\rm HD}$ has been corrected for the recontructed peculiar velocity~\cite{Stiskalek:2025ktq} by Eq.~\eqref{8}. For the sake of consistency, we also adopt the same estimation $\sigma_{v,\rm residual}=172.5\pm27.3\,\mathrm{km\, s^{-1}\,Mpc^{-1}}$ as Ref.~\eqref{eq:app-bias} for the residual small-scale, line-of-sight peculiar-velocity dispersion inferred using the Manticore-Local reconstruction, with their fiducial SN-magnitude selection model and 35 Cepheid hosts. Here, the line-of-sight perculiar velocity $v_{\parallel,i}=v_{\mathrm{Manticore},i}+\boldsymbol V_{\rm ext}\cdot\hat{\boldsymbol n}_i+\epsilon_i$ has distinguish the reconstructed local velocity field $v_{\mathrm{Manticore},i}$ and an additional coherent external flow $\boldsymbol V_{\rm ext}\cdot\hat{\boldsymbol n}_i$ from the remaining stochastic velocity $\epsilon_i\sim\mathcal N(0,\sigma_v^2)$. We report the following value for this covariance-weighted volumetric bias,
\begin{align}\label{eq:app-estimate}
\Delta(-5a_B)=0.060\,\mathrm{mag}.,
\end{align}
which roughly explains half of the observed apparent intercept offset of our forward/backward analysis,
\begin{align}
\Delta(-5a_B)=\begin{cases}
23.932-23.807=0.125\,\mathrm{mag},&\hbox{Forward};\\
23.932-23.802=0.130\,\mathrm{mag},&\hbox{Backward}.
\end{cases}
\end{align}

\subsection{Leave-out analysis}

It is easy to see that three (N4424, N4536, M101) of the host galaxies contribute the most to the estimated 0.06 mag. offset from the volumetric bias. Before carrying out the full hierarchical Bayesian analysis~\cite{Brout:2022vxf,Kenworthy:2022jdh,Stiskalek:2025ktq} in a separate future work, a quick leave-out check can be done by repeating our forward and backward analysis from those second-rung SNe Ia in the remaining 32 host galaxies with respect to the third-rung sample. The results are summarized in Fig.~\ref{fig:leaveoutcheck}.

In the forward check, we obtain
\begin{align}
M_{B} &= -19.383\pm 0.035
\end{align}
from the second rung, and 
\begin{align}
H_{0} &= 68.87^{+1.19}_{-1.15}\,\mathrm{km\,s^{-1}\,Mpc^{-1}}
\end{align}
from the third rung.
For $-5a_{B}$, we obtain
\begin{align}
-5a_{B} &= 23.859 \pm 0.031 \text{ (The second rung)}\\
-5a_{B} &= 23.807 \pm 0.007 \text{ (The third rung)}.
\end{align}

In the backward check, we obtain
\begin{align}
M_{B} &= -19.447\pm 0.015
\end{align}
from the third rung, and 
\begin{align}
H_{0} &= 65.44^{+1.03}_{-1.01}\,\mathrm{km\,s^{-1}\,Mpc^{-1}}
\end{align}
from the second rung.
Also for $-5a_{B}$, we obtain
\begin{align}
-5a_{B} &= 23.858 \pm 0.031 \text{ (The second rung)}\\
-5a_{B} &= 23.802 \pm 0.004 \text{ (The third rung)}.
\end{align}
In both cases, the intercept offsets are $\Delta(-5a_B)=0.052, 0.056$ for the forward and backward checks, respectively, close to the above covariance-weighted volumetric bias~\eqref{eq:app-estimate} with the full sample. Applying the leave-out check brings the second-rung intercept closer to the third-rung intercept than solely considering the peculiar-velocity redshift correction in both forward and backward checks; thus, the significance of the discrepancy between the second- and third-rung intercept $-5a_{B}$ reduces to $1.6\sigma$ for the forward check and $1.8\sigma$ for the backward check, less than half of the original discrepancy.

\section{Conclusions and discussions}\label{sec:conclusion}

In this work, we tested the consistency between the second and third rungs of the distance ladder through forward and backward checks based on the sequential transfer of the $M_B$ calibration. Correcting the host-galaxy redshifts for peculiar velocities using the estimates from Ref.~\cite{Stiskalek:2025ktq} brings the recovered $H_0$ constraints closer to the CMB-inferred value in both analyses, although discrepancies remain. The $-5a_B$ posteriors derived from the two stages provide a complementary diagnostic, allowing us to quantify the significance of the discrepancy between the two rungs before and after the redshift corrections. The significance of the mismatch decreases from $6.8\sigma$ to $3.9\sigma$ in the forward analysis and from $7.3\sigma$ to $4.3\sigma$ in the backward analysis. These results show that the redshift correction partially alleviates the intercept tension, while a residual inconsistency between the two rungs persists. 
The observed alleviation of the calibration intercept $-5a_B$ tension is important in light of the usefulness of using unanchored SNIa luminosity distances in certain cosmological studies (see, e.g., Refs.~\cite{Guo:2024pkx,Liu:2024npr,Liu:2023ifz}).

These results highlight the role of host-galaxy redshift corrections in assessing the consistency of the distance ladder. The shifts in the $H_0$ constraints show that the treatment of peculiar velocities can affect the calibration transferred between the two rungs. Nevertheless, the remaining discrepancies indicate that the adopted corrections do not fully reconcile the two samples within our current analysis framework. Identifying the origin of this mismatch requires further assessment of uncertainties in the peculiar-velocity estimates, sample selection, and possible calibration systematics. For example, as a preliminary check, we can cut the residual intercept offset $\sim0.125$ mag. in half to $\sim0.06$ mag. by weighting all nearby volumetric bias terms and reduce the significance of the residual discrepancy down to $\sim1.7\sigma$ in a leave-out analysis. Comparisons using alternative velocity models and different low-redshift cuts would help establish the robustness of the inferred mismatch from nearby to Hubble-flow ladders. If none of these systematics explains this residual discrepancy, our work can provide a preliminary hint of new physics (e.g. Ref.~\cite{Cai:2021wgv}) operating across the homogeneity scale~\cite{Yu:2022wvg} from the second to the third rungs of the distance ladder.

\begin{acknowledgments}
We thank Adam Riess for insightful suggestions to check for the volumetric bias.
This work was supported by the National Key Research and Development Program of China Grants (No. 2021YFC2203001, No. 2021YFC2203004, and No. 2021YFA0718304), the National Natural Science Foundation of China Grants (No. 12422502, No. 12547110, No. 12588101, No. 12235019, No. 12475046, and No. 12447101), and the program of China Scholarships Council. MB was supported by the Polish National Science Centre grant 2023/50/A/ST9/00579.
\end{acknowledgments}

\bibliography{ref}
\end{document}